\pdfoutput=1
\documentclass[journal]{IEEEtran}
\usepackage{graphicx}
\usepackage[pdftex]{color}
\usepackage[]{amssymb}
\usepackage[]{textgreek}
\usepackage[alsoload=synchem]{siunitx}
\DeclareSIUnit\pixel{pixel}
\DeclareSIUnit\electron{e^{-}}
\DeclareSIUnit\adu{ADU}
\usepackage[cmex10]{mathtools}
\DeclareMathOperator\e{e}
\DeclareMathOperator\W{W}
\usepackage{physics}
\begin{document}
\title{Analytical Solutions of Transient Drift-Diffusion\\in P-N Junction Pixel Sensors}
\author{G.~Blaj*,~\IEEEmembership{Member,~IEEE},
        C.~Kenney,~\IEEEmembership{Member,~IEEE},
        J.~Segal,~\IEEEmembership{Member,~IEEE},
        G.~Haller,~\IEEEmembership{Member,~IEEE}
\thanks{Manuscript received June 5\textsuperscript{th}, 2017; SLAC-PUB-16992.}
\thanks{G.~Blaj, C.~Kenney, J.~Segal and G.~Haller are with SLAC National Accelerator Laboratory, Menlo Park, CA 94025.}
\thanks{* Corresponding author: blaj@slac.stanford.edu}
}
\markboth{}
{Shell \MakeLowercase{\textit{et al.}}: Bare Demo of IEEEtran.cls for Journals}
\maketitle
\begin{abstract}
Radiation detection in applications ranging from high energy physics to medical imaging rely on solid state detectors, often hybrid pixel detectors with (1)~reverse biased \mbox{p-n}  junction pixel sensors and (2)~readout ASICs, attached by flip-chip-bonding. Transient signals characteristics are important in, e.g., matching ASIC and sensor design, modeling and optimizing detector parameters and describing timing and charge sharing properties. Currently analytical forms of transient signals are available for only a few limited cases (e.g., drift or diffusion) or for the steady state (which is not relevant for high energy radiation detection). Tools are available for (relatively slow) numerical evaluation of the transient charge transport. We present here the first analytical solutions of partial differential equations describing drift-diffusion-recombination charge transport in planar \mbox{p-n} junction sensors in a variety of conditions: (1)~undepleted, (2)~fully depleted, (3)~taking into account the gradual velocity saturation, and (4)~overdepleted. We deduce the Green's functions which can be applied to any detection problem through simple convolution with the initial conditions. We compare the analytical solutions with Monte Carlo simulations and industry standard simulations (Synopsys Sentaurus), demonstrating good agreement. Using the analytical equations enables fast modeling of the influence of various detector parameters on tracking, imaging and timing performance, describing performance and enabling optimizations for different applications. Finally, we illustrate this model with applications in 3D+T (x,y,z,time) photon tracking and 4D+T (x,y,\texttheta,\textphi,time) relativistic charged particle tracking. 
\end{abstract}
\IEEEpeerreviewmaketitle

\section{Introduction}

\begin{figure}[!t]
\centering
\includegraphics[width=\columnwidth]{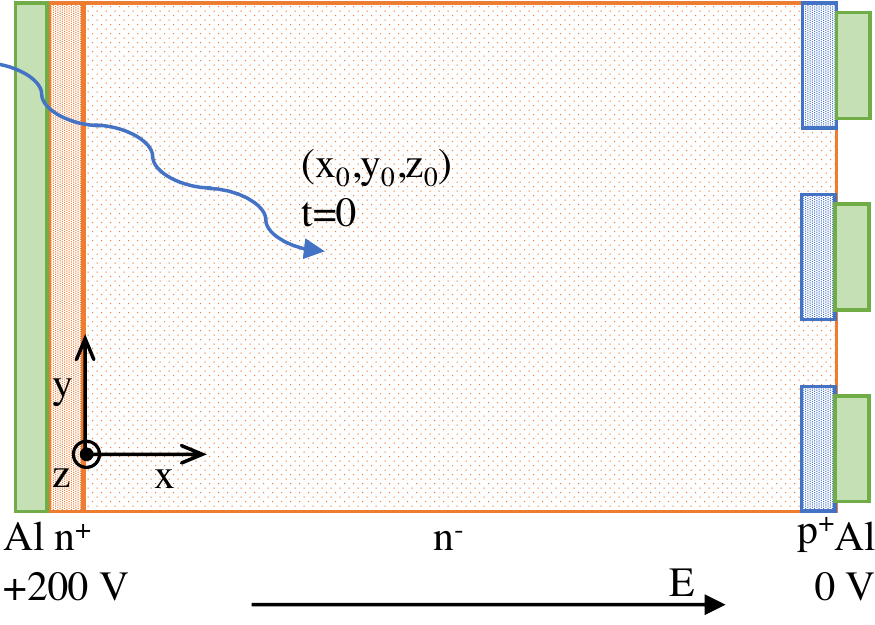}
\caption{Typical n-type sensor for a hybrid pixel detector. The front entrance window Al and n+ region are typically small (less than \SI{1}{\micro\metre} each), and are located at $x \approx 0$. A bias voltage of \SI{200}{\volt} is applied to fully deplete the n- bulk. The rear side contacts are flip-chip-bonded to individual pixel readout channels on a readout ASIC. The minority carriers in the n- bulk are holes. At moment $t=0$ an x-ray photon deposits its entire energy in a concentrated cloud near $(x_0,y_0,z_0)$.}
\label{fig1}
\end{figure}

Radiation detection in applications ranging from high energy physics to medical imaging rely on solid state sensors (often silicon). The last three decades saw significant improvements in hybrid pixel detectors \cite{delpierre2014history}, which allow developing advanced functionality in the CMOS pixel readouts (e.g., digital photon counting \cite{campbell1998readout, llopart2001medipix2,broennimann2006pilatus}, low noise charge integrating \cite{blaj2016xray}, spectroscopy \cite{ballabriga2013medipix3rx, dragone2015epixs}, timing \cite{llopart2007timepix, poikela2014timepix3, markovic2016design}, sparsification \cite{poikela2014timepix3}, gain switching \cite{greiffenberg2012agipd, caragiulo2014design}, other specialized functionality \cite{erdinger2012dssc,philipp2016high}), while separating ASIC and sensor development and leveraging the commercial advances in chip fabrication. Most often, pixel sensors based on reverse biased \mbox{p-n} junctions are used \cite{dalla2015why} (see a schematic diagram in Fig.~\ref{fig1}). Typical sensor materials include Si, GaAs, CdTe, Ge.
Currently the transient signals are analytically described for only a few special cases (e.g., simple thermal diffusion \cite{lutz1999semiconductor}). Steady state solutions are often presented \cite{lutz1999semiconductor, zeghbroeck2011principles}; while they are useful in other applications (imaging at low energy, high flux, saturation, etc.), they are less relevant for discrete high energy quantum detection, where the steady state is trivial. 

Often simulations \cite{plummer2000silicon} with technology computer-aided design (TCAD) packages are used \cite{dutton2000perspectives}, despite being relatively slow and requiring significant training. Other numerical approaches include finite elements: FEMOS \cite{bortolossi20143d}, 2D Monte Carlo: Weightfield2 \cite{cenna2015weightfield2}, simple assumptions about the charge transport: HORUS \cite{potdevin2009horus}, or measurements  and simulations of simple \mbox{p-n} diodes \cite{becker2011measurements} with resulting limitations in speed, ease of use, and/or accuracy. These simulation tools typically require other software frameworks (e.g., IDL, Root) and, with the exception of Weightfield2, are not easily available.

We present here the first analytical solutions of partial differential equations describing drift-diffusion-recombination charge transport in planar \mbox{p-n} junction sensors in a variety of conditions: (1)~undepleted, (2)~fully depleted, (3)~taking into account the gradual velocity saturation, and (4)~overdepleted. We deduce the Green's functions which can be applied to any detection problem through simple convolution with the initial conditions. We compare the analytical solutions with Monte Carlo simulations and industry standard TCAD simulations (Synopsys Sentaurus \cite{synopsys2016sentaurus}), demonstrating good agreement.

Finally, we deduce equations governing transient charge transport and charge sharing, relating detection parameters (location and time, bias voltage, track orientation, and pixel geometry) and providing examples with 3D+T (x,y,z,time) photon tracking and 4D+T (x,y,\texttheta,\textphi,time) relativistic charged particle tracking. These analytical solutions enable fast modeling of the influence of various detector parameters on tracking, imaging and timing performance, describing performance and enabling optimizations for different applications.

\section{Reverse Biased \mbox{p-n} Junction Sensors}

In radiation imaging with semiconductor pixel sensors, detection typically occurs in discrete events, resulting in concentrated electron-hole clouds around points or lines at the location and time of radiation interaction with the semiconductor sensor. These discrete clouds subsequently drift, diffuse and recombine until reaching the highly doped front or rear side contacts (Fig.~\ref{fig1}).

The evolution of signals induced by individual photons or particles are non-equilibrium, non-steady-state processes in which the time evolution of charge carriers is important. 

With high fluxes of radiation, the steady state solution can be useful. However, low noise radiation detection is usually measuring signals from single particles. In this case, the steady state solution is trivial, thus it is necessary to take into account the transient regime (i.e., spatio-temporal evolution of charge carrier concentrations).

\subsection{Sensors}
Intrinsec semiconductors have relatively high thermal noise compared to signals induced by single x-ray photons \cite{lutz1999semiconductor}. To minimize the thermal noise, the sensor material can either be cooled to cryogenic temperatures (e.g., high purity germanium detectors) or used as a reverse biased \mbox{p-n} junction.

Most hybrid pixel sensors used in radiation detection are reverse biased \mbox{p-n} junctions, often a bulk n-type silicon sensor (thicknesses up to \SI{1}{\milli\metre} are common), with a thin ($\le\SI{1}{\micro\metre})$ p implant region. Fig.~\ref{fig1} shows a cross section of a typical n-type silicon sensor, with the front entrance window shown on the left, and several pixel readout contacts on the back plane shown on the right. Other sensor materials (e.g., CdTe, GaAs, Ge) can also be used.

The concentration of dopant $N_D$ in the n-type sensor bulk is related to the resistivity $\rho_n$ by \cite{sze2006physics}:
\begin{align} \label{eq1}
    N_D=\frac{1}{\mu_n q_e \rho_n}
\end{align}
Usually the resistivity is quoted instead of the concentration of dopant. A typical value is $\rho=\SI{10}{\kilo\ohm\centi\metre}$, correspoding to a donor concentration of $N_D\approx\SI{4.34E11}{\per\cubic\centi\metre}$.

\subsection{Bias Voltage and Depletion}
Applying a reverse bias voltage $V$ results in a fully depleted detector (i.e., over the entire detector width $d$), or partially depleted (i.e., depleted over width $W$ close to the readout and undepleted over a width $d-W$ close to the entrance window). We will call the sensor plane near the readout ASIC "rear plane" and the photon entrance plane "front plane". The depletion width can be calculated \cite{zeghbroeck2011principles}:
\begin{align} \label{eq2}
    W=\sqrt{\frac{2 \varepsilon V}{q_e N}}
\end{align}
where $\varepsilon$ is the silicon permittivity, $q_e$ is the elementary charge, $N$ is the dopant density, and $V$ is the bias voltage. 

Table~\ref{table1} summarizes relevant constants for electrons and holes in silicon (values from \cite{becker2011measurements})
\begin{table}[!t]
  \renewcommand{\arraystretch}{1.3}
  \caption{Charge Carriers in Silicon}
  \label{table1}
  \centering
  \begin{tabular}{{l}{l}{l}{l}}
    \hline
     Parameter & $e^-$ & $h^+$ & Units \\
    \hline
    $\mu$ & \num{1440} & \num{474} & \si{\square\centi\metre\per\volt\per\s} \\
    $D$ & \num{36.38} & \num{11.96} & \si{\square\centi\metre\per\second} \\
    $v_s$ & \num{1.054E7} & \num{0.940E7} & \si{\centi\metre\per\second} \\
    \hline
  \end{tabular}
\end{table}

If $W < d$, the sensor is partially depleted. If $W \ge d$, the sensor is fully depleted. The bias voltage $V_D$ to fully deplete a sensor of thickness $d$ results from substituting $W$ and $V$ in Eq.~\ref{eq2} with $d$ and $V_D$:
\begin{align} \label{eq3}
    V_D=\frac{q_e N d^2}{2 \varepsilon}
\end{align}

\subsection{Typical Sensor}
While there are many types of sensors for hybrid pixel detectors, an often used configuration is n-type silicon, with resistivity $\rho=\SI{10}{\kilo\ohm\centi\metre}$ and thickness $d=\SI{300}{\micro\metre}$. A bias of $V_D\approx\SI{30}{\volt}$ will fully deplete such a sensor. Table~\ref{table2} summarizes biasing and drift parameters for the typical sensor.
\begin{table}[!t]
  \renewcommand{\arraystretch}{1.3}
  \caption{Typical silicon sensor}
  \label{table2}
  \centering
  \begin{tabular}{{l}{l}{l}}
    \hline
     Parameter & Value & Formula \\
    \hline
    $\varepsilon_r$ & \num{11.68} & \\
    $\varepsilon$ & \num{1.034E-12}{\farad\per\cm} & $\varepsilon_0 \varepsilon_r$ \\
    $\rho_n$ & \SI{10}{\kilo\ohm\centi\metre} & \\
    $N_D$ & \SI{4.334E11}{\per\cubic\centi\metre} & $1/(\mu_n q_e \rho_n)$\\
    $d$ & \SI{300}{\micro\metre} & \\
    $V_D$ & \SI{30}{\volt} & $q_e N_D d^2/(2 \varepsilon)$ \\
    $W$ & $W(V)$ & $\sqrt{2 \varepsilon V / (q_e N_D)}$ \\
    $b_p$ & \SI{3.178E+7}{\per\second}& $\mu_p / (\mu_n \varepsilon \rho_n)$\\
    $b_n$ & \SI{9.653E+7}{\per\second}& $1/(\varepsilon \rho_n)$\\
    \hline
  \end{tabular}
\end{table}
 
 Throughout this paper we will often refer to and use this "typical sensor" to show examples of how the drift, diffusion and charge sharing would affect detection of photons and relativistic charged particles. However, the equations presented here are generally applicable to any p-n junction sensor material (e.g., Si, CdTe, GaAs, Ge), type (p or n), geometry (sensor thickness, pixels or strips, length and width), with appropriate choices of carriers and integration limits.

\subsection{Electric Field}

\subsubsection{Fully depleted sensor}
The sensor is biased over its entire length $d$. The electric field varies linearly from $E_D=(V-V_D)/d$ at front window to $(V + V_D)/d$ at rear contacts \cite{zeghbroeck2011principles}.

\subsubsection{Partially depleted sensor}
The sensor is depleted over a region of thickness $W$ from the back contacts and undepleted in the remaining volume. The electric field is $E=0$ in the undepleted region and increases linearly to $E=2 V/W$ at the back plane \cite{zeghbroeck2011principles}.

\subsubsection{In general}
\begin{figure}[!t]
\centering
\includegraphics[width=\columnwidth]{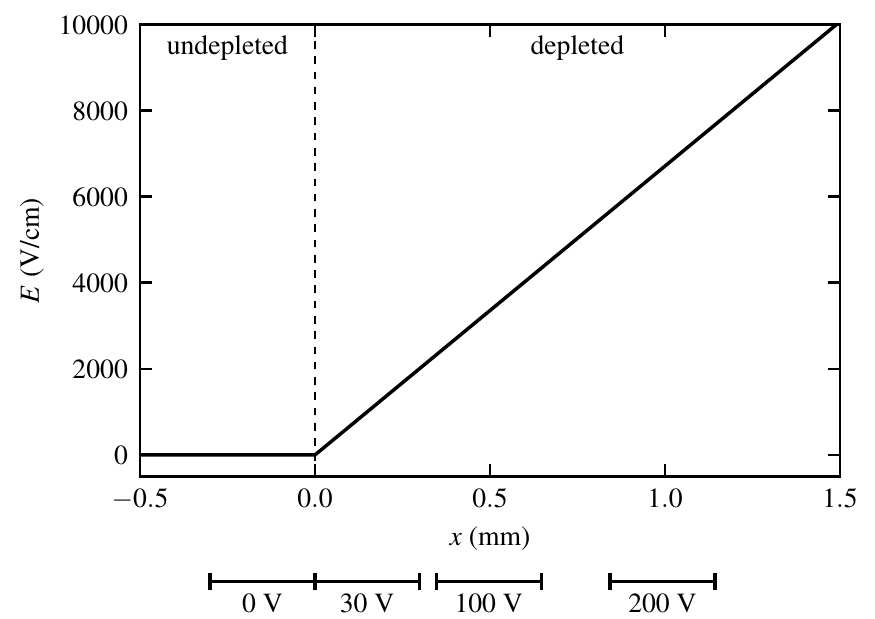}
\caption{Dependence of electric field $E$ on position $x$. We introduce the convention that $x=0$ at the interface between the depleted and undepleted regions, which could be real (inside) or virtual (outside the detector). In this reference system, the front plane is at $x_1(V)$ and the rear plane is at $x_2=x_1+d$. For a few bias conditions $V \in \{0, \SI{30}{\volt}, \SI{100}{\volt}, \SI{200}{\volt}\}$, the $[x_1,x_2]$ coordinates are indicated with labeled line segments under the plot.}
\label{fig2}
\end{figure}

The transport equations are invariant with translation, so we'll conveniently set the origin of $x$ axis at the interface between the depleted and undepleted regions, which could be real (inside) or virtual (outside the detector). This choice of coordinate system origin will greatly simplify accounting for offsets in subsequent sections. The resulting electric field is:
\begin{align} \label{eq4}
    E(x)=\frac{2 V_D}{d^2} x = \frac{1}{\varepsilon \mu_n \rho_n} x
\end{align}
Fig.~\ref{fig2} shows the electric field $E(x)$ dependence on position and illustrates detector coordinates $[x_1,x_2]$ for a few different bias voltages.

\subsection{Arbitrary Bias Voltage}
With the $x$ origin at the interface between the undepleted and depleted regions, different biasing conditions result in different coordinates of the front and rear planes of the sensor (indicated in Fig.~\ref{fig2} for $V\in\{0,\SI{30}{\volt},\SI{100}{\volt},\SI{200}{\volt}\}$).

The voltage drop over the sensor can be obtained by integrating Eq.~\ref{eq4} over $x$ (for simplicity we discard signs here); $x_1$ and $x_2$ are the positions of the front and back of the sensor, with $x_2-x_1=d$. In a partially depleted sensor, $x_2=W$ thus $x_1=W-d$. In a fully depleted sensor, the bias voltage is $V=\frac{1}{\varepsilon \mu_n \rho_n} d \left( x_1 + \frac{d}{2}\right)$ resulting in:
\begin{align}
x_1= \label{eq5}
\begin{cases}
    W-d & \text{if $W\le d$}\\
    \frac{\varepsilon \mu_n \rho_n V}{d} - \frac{d}{2} & \text{if $W>d$}
\end{cases}\\
x_2=x_1+d \label{eq6}
\end{align}

\subsection{Charge Generation, Transport and Recombination}
Detecting individual quanta in semiconductors relies on the transient signals induced by charge transport (drift and diffusion), which in turn depends on charge generation at the interaction point(s) and recombination of carriers.

\subsubsection{Charge Generation}
Different particles or photons generate distinct patterns that can be used to identify the detected particle. For example, visible and UV light photons generate single electron pairs. High fluxes are adequately described by existing steady state approximations.

X-ray photons generate discrete electron-hole clouds with thousands of carriers near the interaction point. Relativistic charged particles typically pass through the sensor in straight lines, depositing a constant amount of energy per length $\dv{E}{x}$ (Bethe formula). The signal can be integrated as a series of small signals over the track length. The steady state solution for x-ray photons and relativistic particles is trivial, and the transient signals have to be evaluated.

\subsubsection{Thermal Diffusion}
Charge carriers undergo thermal diffusion, with an initial $\delta(x)$ distribution of carriers at $t=0$ (in the absence of electric fields or boundary conditions) diffusing to a normal distribution with size \cite{logan2014applied}:
\begin{align} \label{eq7}
    \sigma=\sqrt{2 D t}
\end{align}
where the diffusion constant $D$ is given by the Einstein relation \cite{einstein1905molekularkinetischen}: $D=\mu k_B T / q_e$.

\subsubsection{Charge Drift}
Charge drift is determined by electric fields (accelerating charge carriers) and interactions with the lattice. At high field intensities, the average drift velocity asymptotically approaches the saturation velocity $v_s$ as the interactions with the lattice balance out the acceleration in the electric field. For $x>0$, $E$ induces a carrier velocity component (drift velocity), which for indirect band gap semiconductors can be written as:
\begin{align}
    v(x)=\frac{\mu E(x)}{1+\frac{\mu E(x)}{v_s}}=\frac{b x }{1 + \frac{b x}{v_s}} =\frac{b x v_s}{b x + v_s} \text{, with} \label{eq8}\\
    b_n=\frac{1}{\varepsilon \rho_n} \text{, } b = b_p=\frac{\mu_p}{\mu_n \varepsilon \rho_n}  \label{eq9}
\end{align}
where $v_s$ is the saturation velocity. We call this a "saturation velocity model".

We introduce the $b_p$ and $b_n$ constants (Eq.~\ref{eq9}), as they will be used extensively throughout this paper. For simplicity we will often use $b$ instead of $b_p$. They are constant for each sensor (depending only on doping and, for minority carriers, also on mobility). Note that the forms in Eq.~\ref{eq9} are valid for n-type sensors (i.e., electrons are majority carriers and holes are minority carriers). The $b$ coefficients could be called "linear velocity gradients".

\begin{figure}[!t]
\centering
\includegraphics[width=\columnwidth]{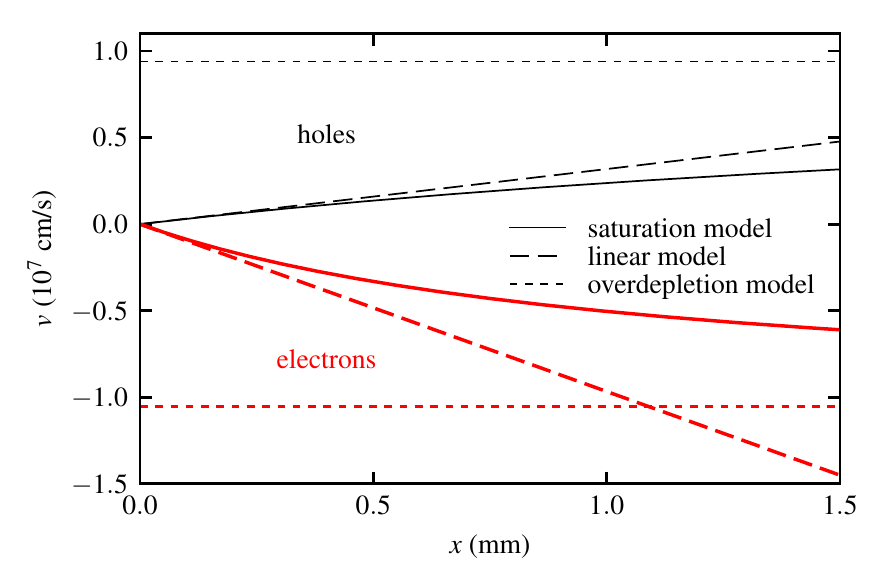}
\caption{Charge carrier velocity in Si as function of position-dependent electric field. Thin black lines (above zero) correspond to holes and thick red lines (below zero) correspond to electrons. Solid lines depict the saturation model (Eq.~\ref{eq8}), dashed lines show to the linear model (Eq.~{\ref{eq10}}) while the dotted lines show the overdepletion model (Eq.~\ref{eq11}).}
\label{fig3}
\end{figure}

We show an example of drift velocity in Fig.~\ref{fig3} for minority carriers (holes, black lines) and majority carriers (electrons, red lines) in a typical sensor. Solid lines correspond to the saturation velocity model (Eq.~\ref{eq8}) while dashed lines correspond to the linear velocity model (Eq.~\ref{eq10}), demonstrating significant differences at bias voltages of \SI{200}{\volt} in a typical sensor.

When ignoring the saturation effect of carrier velocity ($v_s \gg v(x)$), Eq.~\ref{eq8} is simplified to a "linear velocity model":
\begin{align} \label{eq10}
    \lim_{v_s \to \infty} v(x)=\mu E(x) = b x
\end{align}
which is also appropriate for describing carrier drift in direct band gap semiconductors under the peak velocity.

If the sensor is overdepleted (i.e., $V \gg V_D$), the carrier velocity approaches the saturation velocity asymptotically:
\begin{align} \label{eq11}
    \lim_{V \to \infty} v(x) = v_s
\end{align}
further called "overdepletion model".

\subsubsection{Charge Recombination}
Charge recombination typically has a time constant in the order of milliseconds \cite{zeghbroeck2011principles}, much larger than drift times (typically in the order of nanoseconds, Table~\ref{table3}) and can usually be ignored. If this is not the case (e.g., after significant radiation damage \cite{loferski1958radiation}), the appropriate recombination rate $\gamma$ can be used in the full form solutions (Eq.~\ref{eq13},~\ref{eq42}).

\subsection{Partial Differential Equations}
Solving the drift-diffusion-recombination partial differential equation yields the transient signals from single detected quanta. We solve and discuss the partial differential equations for minority carriers (i.e., holes in n-type sensors) as they induce most of the signal into the pixel readout. We will briefly mention the majority carriers and their solutions.

\subsubsection{General Equation}
The hole density as a function of time and space can be written as $p(x,y,z,t)$. The effects of charge transport and recombination mechanisms can be summarized as:
\begin{align} \label{eq12}
    \pdv{p}{t}=D \grad^2 p - \grad (v(x) p) - \gamma p
\end{align}
where the right hand side terms correspond to thermal diffusion, drift in the electric field\footnote{The minus sign for the drift term in Eq.~\ref{eq12} is correct for positive $v$ equivalent to moving to the right.}, and charge recombination, respectively. A similar equation is valid for electron density $n(x,y,z,t)$.

While these two equations are coupled and electrostatic effects are present \cite{segal1996simulation}, in both the depleted and undepleted sensor the coupling is relatively weak. Electrostatic interactions become important when large signals are detected in a small volume and short time, resulting in "plasma effects" \cite{becker2010simulation}.

\subsubsection{Field Along x Axis}
The electric field components in the $yz$ plane can usually be neglected\footnote{Close to the rear pixel contacts there are electric field components in-plane, however, their influence is relatively small as minority carriers drift relatively quickly through this region and are unlikely to diffuse to nearby pixels.}. Usually the recombination rate does not depend on position and time, allowing us to extract the recombination term $-\gamma p$ and multiply the solution with a factor $\e^{-\gamma t}$ instead. This allows separating the variables
\begin{align} \label{eq13}
p(x,y,z,t) = u(x,t) w(y,z,t) \e^{-\gamma t}
\end{align}
and results in a 1D drift-diffusion equation (also called diffusion-advection) along the $x$ axis and diffusion in the $yz$ plane. Thus we have to solve Eq.~\ref{eq12} only in one dimension for $u(x,t)$:
\begin{align}
    \pdv{u}{t}=D \pdv[2]{u}{x} - \pdv{(v(x) u)}{x}  \label{eq14}\\
    u(x,0)=\delta(x-\xi_0) \label{eq15}
\end{align}

\subsubsection{Charge generation}
For clarity, and in line with the prevailing notation (e.g., \cite{logan2014applied}), we will denote the initial conditions on  $x$, $y$, and $z$ axes with $\xi$, $\upsilon$ and $\zeta$. A discrete detection event at $t=0$ and location $(\xi_0,\upsilon_0,\zeta_0)$ will generate a charge cloud:
\begin{align} \label{eq16}
    p(x,y,z,0)=\delta(\xi-\xi_0)\delta(\upsilon-\upsilon_0)\delta(\zeta-\zeta_0)
\end{align}

Solving the carrier density equation for infinitely small initial distributions (i.e., $\delta$ functions) yields the Green's function corresponding to the partial differential equation and initial and boundary conditions. For finite initial signals, the solution is a simple convolution of the Green's function with the initial signal \cite{logan2014applied}. For $t < 0$, $p(x,y,z,t)$, $u(x,t)$, and $w(y,z,t)$ are zero.

\subsubsection{Lateral Charge Diffusion}
Assuming the sensor is very large in the $yz$ plane and ignoring in-plane electric field components, charge drift as a function of time $w(y,z,t)$ will be determined by:
\begin{align}
    \pdv{w}{t}= D \grad^2 w \label{eq17} \\
    w(y,z,0) = \delta(\upsilon-\upsilon_0) \delta(\zeta-\zeta_0) \label{eq18}
\end{align}
with the familiar 2D diffusion solution:
\begin{align}
    w(y,z,t)=\frac{1}{4 \pi D t}\exp(-\frac{(y-\upsilon_0)^2}{4 D t}) \exp(-\frac{(z-\zeta_0)^2}{4 D t}) \label{eq19}
\end{align}

\section{Overdepletion Velocity Model}
\subsection{Green's Function}
With constant velocity $v=v_s$ (Eq.~\ref{eq11}), this is essentially the diffusion model, drifting with constant velocity $v_s$. The corresponding drift equation is:
\begin{align} \label{eq20}
    x_c(\xi,t)=\xi+v_s t
\end{align}
and diffusion equation along $x$ (due to the absence of a velocity gradient, it reverts to simple thermal diffusion):
\begin{align} \label{eq21}
    \sigma(t)=\sqrt{2 D t}
\end{align}
resulting in a relatively simple Green's function for minority carriers:
\begin{align} \label{eq22}
    g(x,\xi,t)=\frac{\exp(-\frac{(x-\xi-v_s t)^2}{4 D t})}{\sqrt{4 \pi D t}}
\end{align}

\section{Linear Velocity Model \label{sec3}}
In the linear velocity approximation we solve the partial differential equation (Eq.~\ref{eq14}) with initial conditions Eq.~\ref{eq15} and $v(x) = b x$ (Eq.~\ref{eq10}). At the boundaries, the recombination is instantaneous, thus $u(x_1,t)=0$ and $u(x_2,t)=0$. Usually drift dominates diffusion at the boundaries so we can ignore the boundary conditions (see section~\ref{sec4E} for a discussion on when this approximation is appropriate).

\subsection{Drift\label{sec3a}}
Single charge carriers drift and diffuse randomly. In localized clouds composed of many carriers, the drift of the center of the cloud will average out the stochastic diffusion of individual charge carriers, drifting from $\xi \in [x_1,x_2]$ (inside the sensor) to $x \in [\xi,x_2]$ in a time $t_d$:
\begin{align} \label{eq23}
    t_d(\xi,x)=\int_{\xi}^{x}\frac{dx}{v(x)} =\frac{1}{b}\ln(\frac{x}{\xi })
\end{align}
Solving for $x$ yields the charge cloud position $x_c(\xi,t)$:
\begin{align} \label{eq24}
    x_{c}^p(\xi,t)=\xi \e^{b t}
\end{align} 

The drift equation for majority carriers is obtained similarly:
\begin{align} \label{eq25}
    x_{c}^n(\xi,t)=\xi \e^{-b t}
\end{align}

\subsection{Diffusion \label{sec3b}}
In appendices~\ref{appendix1}~and~\ref{appendix2} we present an approach to separate diffusion from drift (similar to the method of characteristics \cite{logan2014applied}), reducing the partial differential equation (Eq.~\ref{eq14}) to an ordinary differential equation (Eq.~\ref{eq56}) and obtaining the analytical solution for diffusion of minority carriers:
\begin{align} \label{eq26}
    \sigma_p(t)=\sqrt{\frac{D}{b} \left(\e^{2 b t}-1\right)}
\end{align}
Note that $\lim_{b \to 0}\sigma(t)=\sqrt{2 D t}$ (calculated using \cite{wolfram2017mathematica}) thus Eq.~\ref{eq26} is a generalized form of the simple diffusion equation (Eq.~\ref{eq7}), incorporating the linear velocity gradient $b$.

Similarly for majority carriers, the diffusion equation is:
\begin{align} \label{eq27}
    \sigma_n(t)=\sqrt{\frac{D}{b} \left(1-\e^{- 2 b t}\right)}
\end{align}

\subsection{Green's Function}
With $x_{cp}(\xi,t)$ and $\sigma_p(t)$ calculated above (Eq.~\ref{eq24},~\ref{eq26}), the Green's function for the linear velocity model is:
\begin{align} \label{eq28}
    g(x,\xi,t)=\frac{\exp \left(-\frac{ \left(x-\xi \e^{b t}\right)^2}{2 \frac{D}{b} (\e^{2 b t}-1)}\right)}{\sqrt{2 \pi} \sqrt{\frac{D}{b} \left(\e^{2 b t}-1\right)}}
\end{align}

The Green's function can be used to calculate $u(x,t)$ for any initial condition $u(\xi,0)=f(\xi)$ through convolution over the initial conditions:
\begin{align} \label{eq29}
    u(x,t)=\int_{x_1}^{x_2} g(x,\xi,t) f(\xi) d\xi
\end{align}
For initial condition $f(\xi)=\delta(\xi-\xi_0)$ (point source at $\xi=\xi_0$ and $t=0$), the integral above is reduced to the simple form:
\begin{align} \label{eq30}
    u(x,t)=g(x,\xi_0,t)
\end{align}

\subsection{Testing and Simulations \label{sec3D}}
Substituting the Green's function from Eq.~\ref{eq28} and drift velocity from Eq.~\ref{eq10} in the partial differential equation Eq.~\ref{eq14}, calculating the partial derivatives and canceling identical terms demonstrates that Eq.~\ref{eq28} is the analytical solution.

\begin{figure}[!t]
\centering
\includegraphics[width=\columnwidth]{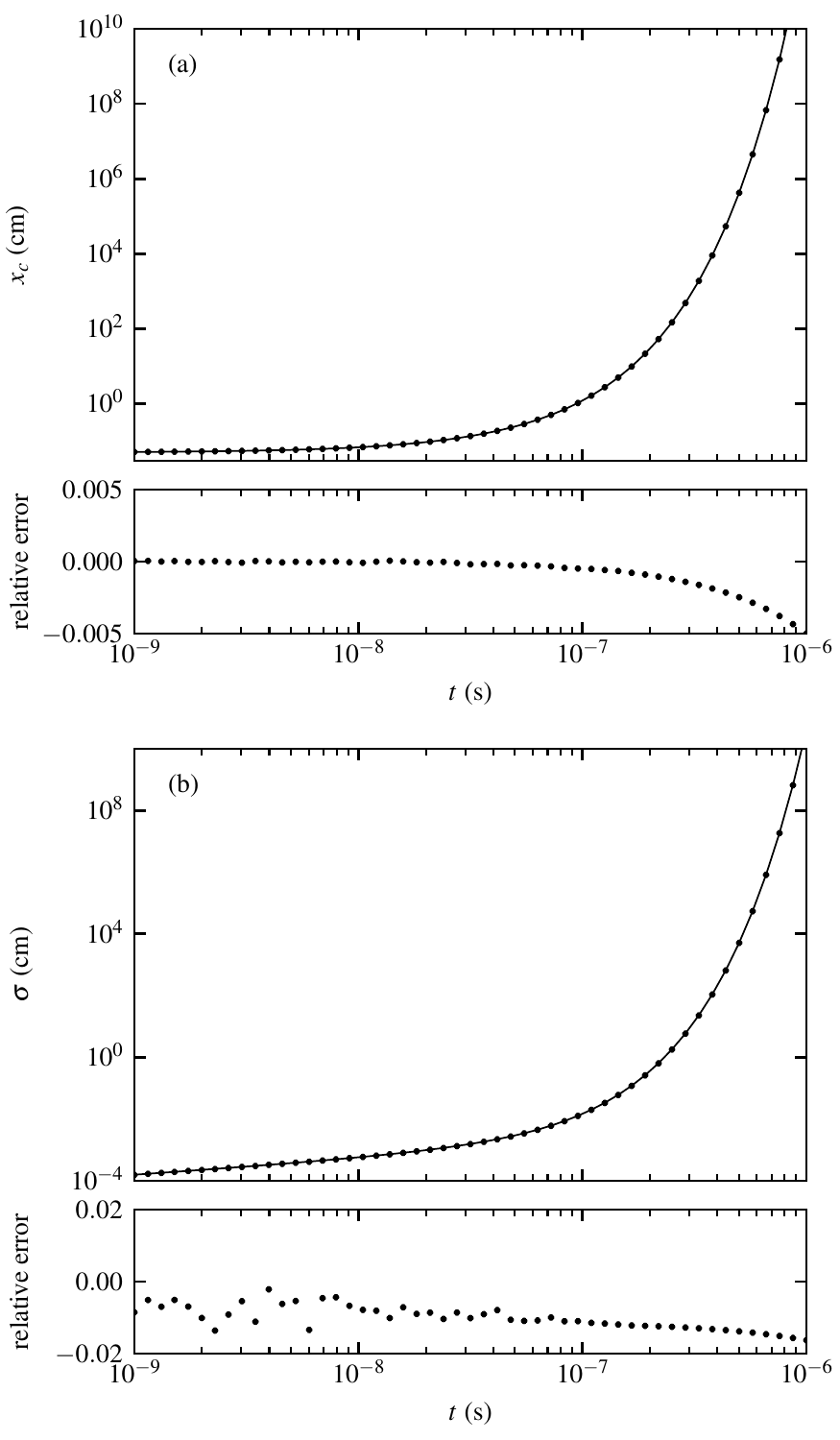}

\caption{Analytical functions (lines) and Monte Carlo simulations (dots) for holes using the linear velocity model of charge transport in an infinitely long sensor (resistivity $\rho_n=\SI{10}{\kilo\ohm\centi\metre}$, initial position $\xi=\SI{0.05}{\centi\metre}$). Lower subplots indicate the relative error. (a) shows the position of the charge cloud as a function of time, with relative errors less than \num{5e-3}; (b) shows the charge cloud size in the x direction, reflecting the supplemental stretching due to the velocity gradient; relative errors less than \num{2e-2}. There is good agreement between the analytical solution and the simulation. Note however that the linear velocity model yields nonphysical results for longer times and large detectors.}
\label{fig4}
\end{figure}

To confirm the analytical results, we performed Monte Carlo simulations (tracking \num{e+5} carriers in a typical sensor, with initial position $\xi_0=\SI{500}{\micro\metre}$) and found that a time step $\Delta t=\SI{10}{\pico\second}$ yields stable results. We simulated a relatively long time ($\SI{1}{\micro\second}$) to prove the validity of the analytical solutions over wide ranges. The results of the numerical simulation are compared to the analytical functions in Fig.~\ref{fig4}, showing a good fit between the analytical model and simulations.

However, the results are obviously nonphysical for large drift times and sensor thicknesses due to the exponential velocity increase implied by Eq.~\ref{eq24}. We will account for the saturation velocity in a generalized saturation velocity model in section~\ref{sec4}, comparing the two models and discussing when to use each (section~\ref{sec4f}).

\section{Saturation Velocity Model \label{sec4}}
With larger bias voltages $V > V_D$, the saturation velocity is usually important, thus we'll solve the partial differential equation Eq.~\ref{eq14} with initial conditions in Eq.~\ref{eq15} and the saturation velocity model in Eq.~\ref{eq8}. As in section~\ref{sec3}, we ignore the boundary conditions (see section~\ref{sec4E} for a discussion on when this approximation is appropriate).

\subsection{Drift}
Following the approach in section~\ref{sec3a}, the drift time for minority carriers from $\xi$ to $x$ is:
\begin{align} \label{eq31}
    t_d^p(\xi,x)=\int_{\xi}^x \frac{dx}{v(x)}=\frac{x-\xi}{v_s}+\frac{\ln (\frac{x}{\xi})}{b}
\end{align}
Solving for $x$, we obtain the drift equation for minority carriers $x_{cp}$:
\begin{align} \label{eq32}
    x_c^p(\xi,t)=\frac{v_s}{b} \W\left(\frac{b \xi}{v_s} \e^{b t +\frac{b \xi}{v_s}}\right)
\end{align}
where $\W$ is the Lambert W function. The drift time and drift equation for majority carriers are obtained similarly:
\begin{align} \label{eq33}  
    t_d^n(\xi,x)=\int_{\xi}^x \frac{dx}{-v(x)}=\frac{\xi-x}{v_s}+\frac{\ln (\frac{\xi}{x})}{b}\\
    x_c^n(\xi,t)=\frac{v_s}{b_n} \W\left(\frac{b_n \xi}{v_s} \e^{- b_n t +\frac{b_n \xi}{v_s}}\right) \label{eq34}  
\end{align}

\subsection{Diffusion}
Similarly to section~\ref{sec3b}, in appendices~\ref{appendix1}~and~\ref{appendix3} we obtain an ordinary differential equation for diffusion (Eq.~\ref{eq61}), resulting in the diffusion equation for minority carriers:
\begin{multline} \label{eq35}
    \sigma_p(\xi,t)=\\
    \frac{\sqrt{
    \frac{b D x_c^2}{v_s^2} \left(2 b t + 4 \ln(\frac{x_c}{\xi})\right)
    + \frac{6 D x_c}{v_s}(\frac{x_c}{\xi}-1) +\frac{D}{b}(\frac{x_c^2}{\xi^2}-1)
    }}{1+\frac{b x_c}{v_s}}
\end{multline}
with $x_c(\xi,t)$ given by Eq.~\ref{eq32}. Note that $\sigma$ depends on initial position $\xi$. In the limit $v_s \to \infty$, this equation simplifies~\cite{wolfram2017mathematica} to Eq.~\ref{eq26}:
\begin{align} \label{eq36}
    \lim_{v_s \to \infty} \sigma_p(\xi,t) = \sqrt{\frac{D}{b}(\e^{2 b t}-1)}
\end{align}
demonstrating that Eq.~\ref{eq35} is a further generalization of Eq.~\ref{eq26}, incorporating the effects of velocity saturation.

Similarly for majority carriers (moving in the opposite direction), with $x_c(\xi,t)$ given by Eq.~\ref{eq34}:
\begin{multline} \label{eq37}
    \sigma_n(\xi,t)=\\
    \frac{\sqrt{
    \frac{b_n D x_c^2}{v_s^2} \left(2 b_n t + 4 \ln(\frac{\xi}{x_c})\right)
    + \frac{6 D x_c}{v_s}(1-\frac{x_c}{\xi}) +\frac{D}{b_n}(1-\frac{x_c^2}{\xi^2})
    }}{1+\frac{b_n x_c}{v_s}}
\end{multline}

\subsection{Green's Function}
With $x_c(\xi,t)$ from Eq.~\ref{eq32} and $\sigma(\xi,t)$ from Eq.~\ref{eq35}, the Green's function $g(x,\xi,t)$ for minority carriers in the saturation velocity model is:
\begin{align} \label{eq38}
    g(x,\xi,t)=\frac{\exp(-\frac{(x-x_c)^2}{2 \sigma^2 })}{\sqrt{2 \pi} \sigma}
\end{align}
with the appropriate $x_c$ and $\sigma$ from Eq.~\ref{eq32} and \ref{eq34} and similarly for the majority carriers, using Eq.~\ref{eq35} and \ref{eq37}.

\subsection{Simulations}
\begin{figure}[!t]
\centering
\includegraphics[width=\columnwidth]{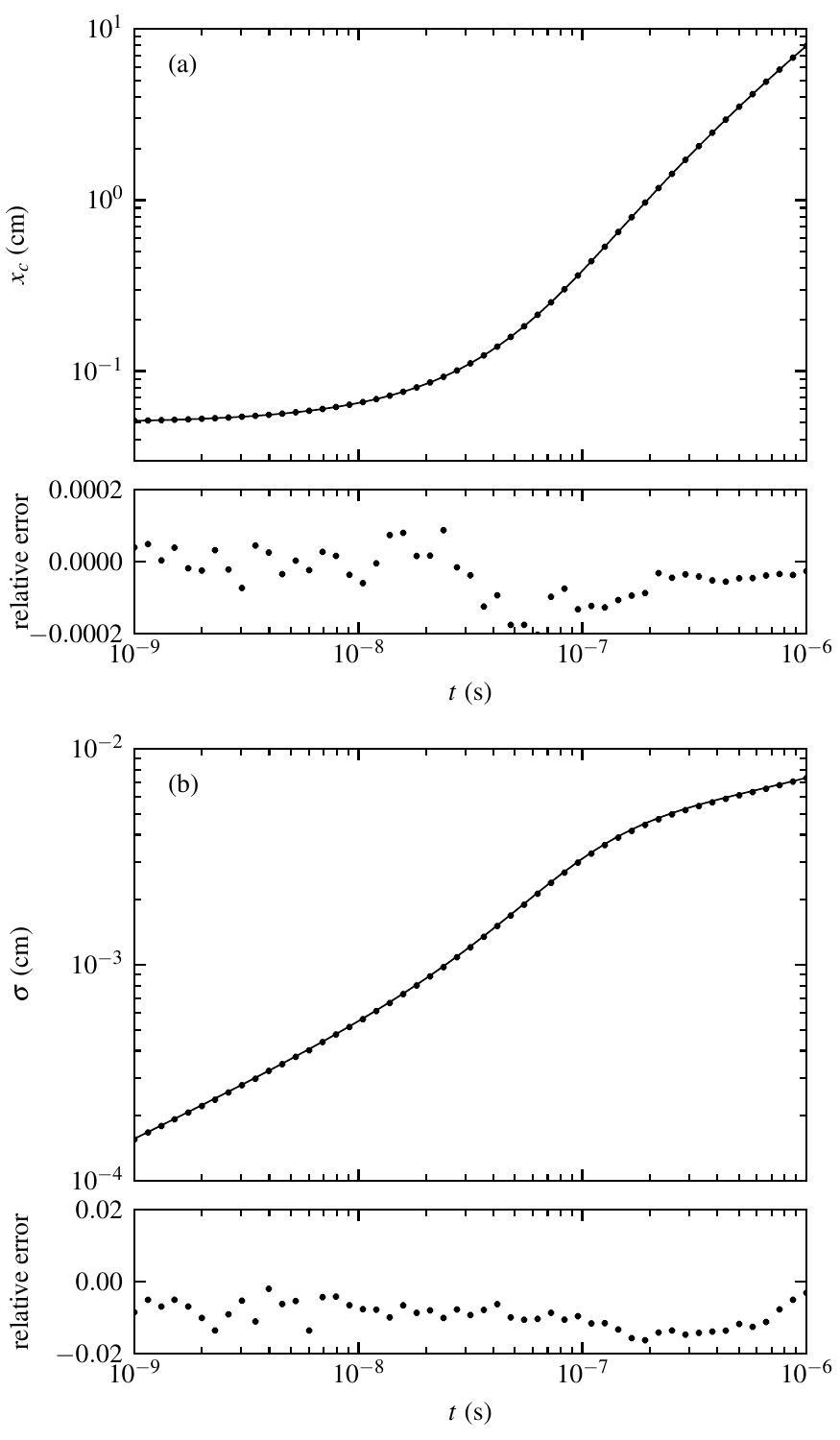}

\caption{Analytical functions (lines) and Monte Carlo simulations (dots) for holes using the saturation velocity model of charge transport in an infinitely long sensor (resistivity $\rho_n=\SI{10}{\kilo\ohm\centi\metre}$, initial position $\xi=\SI{0.05}{\centi\metre}$). Subplots indicate the relative error. (a) shows the position of the charge cloud as a function of time; errors less than \num{E-4}; (b) shows the charge cloud size in the x direction, reflecting the supplemental stretching due to the velocity gradient; errors less than \num{E-2}.}
\label{fig5}
\end{figure}

We performed Monte Carlo simulations as described in section~\ref{sec3D}. Numerical simulations confirm the charge cloud is close to a normal distribution (with negligible skewness and kurtosis), drifting along the $x$ axis. Fig.~\ref{fig5} shows the simulation results (dots) and analytical functions (lines) for both charge cloud position $x_c$ and size (along $x$ axis) $\sigma$, demonstrating good agreement between the analytical model and the simulations (relative error within \num{2e-4} and \num{2e-2} for the position and size, respectively). Other initial positions $\xi \ge 0$ also result in a good match.

\subsection{Boundary Conditions \label{sec4E}}
The Green's function in Eq.~\ref{eq38} is valid for relatively large bias voltages where drift dominates diffusion at both the front and back planes. At low bias voltages, some charge is lost on recombination on the front surface. The charge loss fraction due to recombination on the front plane can be shown to be smaller than or equal to the least advantegeous case (small $\xi_0$, large $t$, carrier transport dominated by diffusion):
\begin{multline} \label{eq39}
    f_{loss}(\xi_0) \le \lim_{t \to \infty} \lim_{x_1 \to 0} \lim_{\xi_0 \to x_1} \int_{-\infty}^{x_1} g(x,\xi_0,t) dx \\
    = \left[1 - \erf \left(-\frac{x_1 \sqrt{b}}{\sqrt{2 D}}\right)\right]/2
\end{multline}
obtained by substituting $g(x,\xi,t)$ from Eq.~\ref{eq28}, which is appropriate for small $\xi_0$.

For a detector with a signal to noise ratio of \num{1000} and a typical sensor, Eq.~\ref{eq28}~and~\ref{eq38} can be used directly for an initial position with a charge loss fraction smaller than \num{e-3}. For the typical sensor we obtain $x_1 \ge \SI{19}{\micro\metre}$, which can be guaranteed with a depletion width $W \ge \SI{319}{\micro\metre}$, corresponding to a bias voltage $V\ge\SI{34}{\volt}$ (obtained using Eq.~\ref{eq3}).

For $V \in [\SI{30}{\volt},\SI{34}{\volt}]$, the partial differential equation Eq.~\ref{eq14} becomes bounded on a semi-infinite domain. One can account for the front boundary condition in this case by adding a virtual negative source, mirrored by the front surface, as shown in chapter~{2.4} of \cite{logan2014applied}.

\subsection{Comparison with Overdepletion Model}
The overdepletion approximation results in large errors for drift in a typical sensor, see summary in Table~\ref{table3}, with errors up to a factor \num{14} for a typical sensor. At \SI{200}{\volt} bias in the typical sensor, the diffusion equation error is up to $\approx\SI{55}{\percent}$, as shown in Fig.~\ref{fig6}~(b) and (c). This approximation is appropriate only for thin p-type sensors with high bias voltages.
\begin{table}[!t]
  \renewcommand{\arraystretch}{1.3}
  \caption{Front to back drift time for minority carriers}
  \label{table3}
  \centering
  \begin{tabular}{{c}{c}{c}{c}{c}}
    \hline
    sensor & $V (\si{\volt})$ & $t_d (\si{\nano\second})$ & $t_d^o (\si{\nano\second})$ & $t_d^l(\si{\nano\second})$\\
    \hline
    n-type, \SI{300}{\micro\metre} & \num{50} & \num{47.17} & \num{3.19} & \num{43.98}\\
    n-type, \SI{300}{\micro\metre} & \num{100} & \num{22.79} & \num{3.19} & \num{19.60}\\
    n-type, \SI{300}{\micro\metre} & \num{200} & \num{12.76} & \num{3.19} & \num{9.57}\\
    \hline
    p-type, \SI{300}{\micro\metre} & \num{100} & \num{13.65} & \num{2.85} & \num{10.80}\\
    p-type, \SI{300}{\micro\metre} & \num{200} & \num{6.70} & \num{2.85} & \num{3.85}\\
    p-type, \SI{300}{\micro\metre} & \num{400} & \num{5.07} & \num{2.85} & \num{2.22}\\
    \hline
    n-type, \SI{75}{\micro\metre} & \num{50} & \num{3.17} & \num{0.80} & \num{2.37}\\
    n-type, \SI{75}{\micro\metre} & \num{100} & \num{1.98} & \num{0.80} & \num{1.19}\\
    n-type, \SI{75}{\micro\metre} & \num{200} & \num{1.39} & \num{0.80} & \num{0.59}\\
    \hline
    p-type, \SI{75}{\micro\metre} & \num{50} & \num{2.12} & \num{0.71} & \num{1.41}\\
    p-type, \SI{75}{\micro\metre} & \num{100} & \num{1.64} & \num{0.71} & \num{0.93}\\
    p-type, \SI{75}{\micro\metre} & \num{200} & \num{1.34} & \num{0.71} & \num{0.63}\\
    \hline
    \multicolumn{5}{l}{$t_d$ is the drift time from the front to the back of the sensor (Eq.~\ref{eq31},~\ref{eq33});}\\
    \multicolumn{5}{l}{$t_d^l$ is calculated in the linear velocity model (Eq.~\ref{eq24},~\ref{eq25});} \\
    \multicolumn{5}{l}{$t_d^o$ is calculated in the overdepletion approximation as $d/v_s$.} \\
  \end{tabular}
\end{table}
\subsection{Comparison with Linear Velocity Model\label{sec4f}}
\begin{figure}[!t]
\centering
\includegraphics[width=\columnwidth]{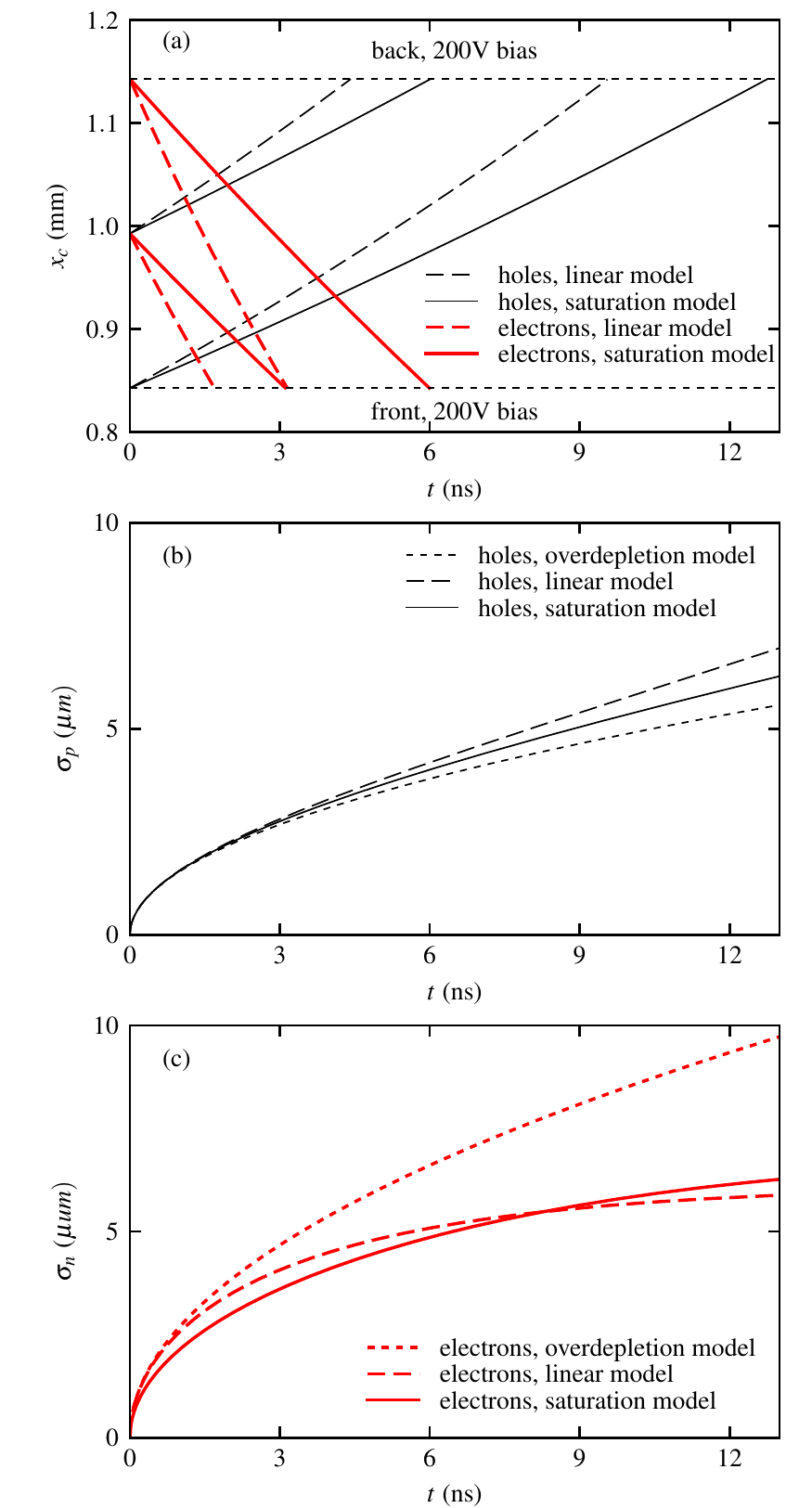}

\caption{Comparing the saturation velocity model (solid lines) with the linear velocity model (dashed lines) and simple diffusion (dotted lines); (a) shows the position of the charge cloud as a function of time; black thin lines (moving up) correspond to holes and red thick lines (moving down) correspond to electrons. The linear velocity model introduces errors of up to \SI{25}{\percent} compared to the saturation velocity model within the typical sensor. (b) shows the cloud charge $\sigma$ as a function of time for minority carriers (holes). The dotted line shows a simple model ignoring the velocity gradient. The simple model underestimates the actual cloud charge by \SI{10}{\percent} while the linear model overestimates it by \SI{10}{\percent}. (c) shows the cloud charge $\sigma$ for majority carriers (electrons). The linear model largely agrees with the saturation model (errors within \SI{5}{\percent}) while the simple model overestimates it by up to \SI{55}{\percent}.}
\label{fig6}
\end{figure}

The saturation velocity model describes accurately the transient charge transfer in thick detectors or detectors with large bias voltages. To investigate the differences between the linear velocity model and the saturation velocity model, we calculated the drift and diffusion for a typical sensor, showing the results in Fig.~\ref{fig6}.

Both minority carriers (thin black lines) and majority carriers (thick red lines) were tracked through the detector volume, with a bias voltage $V=\SI{200}{\volt}$. Three models were used: saturation drift and diffusion (solid lines), linear drift and diffusion (dashed lines) and simple diffusion ignoring the velocity gradients (dotted lines). We used three initial positions: on the front plane, middle of detector, and rear plane.

The three drift models diverge quickly, yielding differences in arrival times of up to \SI{20}{\percent} for minority carriers and up to \SI{40}{\percent} for majority carriers. The diffusion equations for minority carriers diverge more gradually, with similar results in the first nanoseconds. Only the linear and saturation diffusion equations for majority carriers are similar over a wide range of time (due to the thermal drift balancing the compressing velocity gradient).

Consequently, the saturation velocity model should always be used for indirect band gap semiconductors, unless a thin sensor with low bias $V\approx V_D$ is used. However, see section~\ref{sec4E} for limitations associated with bias voltages close to the depletion voltage.

For direct band gap semiconductors (e.g., GaAs, CdTe), the carrier velocity profile as a function of the electric field increases more linearly up to a peak velocity $v_p$ at $E_p$ and then decreases asymptotically to a saturation velocity $v_s$. In this case, the linear model should be used up to $E_p$, and possibly extended piecewise with the overdepleted model for high electric fields.

\section{Undepleted Sensor}
In the case of undepleted sensors we must take into account the boundary conditions $u(-d,t)=0$, $u(0,t)=0$ and the charge recombination rate $\gamma$. We will assume that the sensor is depleted over at least a shallow width, to prevent high thermal noise. The partial differential equation Eq.~\ref{eq14} is reduced to a standard Dirichlet problem \cite{logan2014applied} with corresponding Green's function expressed as an infinite sum:
\begin{align} \label{eq40}
    g(x,\xi,t) = \frac{2}{U} \e^{-\gamma t} \sum_{n=1}^{\infty} \e^{-\frac{D \pi^2 n^2 t}{U^2} }
   \sin \left(\frac{n \pi  x}{U}\right) \sin \left(\frac{n \pi  \xi }{U}\right)
\end{align}

\begin{figure}[!t]
\centering
\includegraphics[width=\columnwidth]{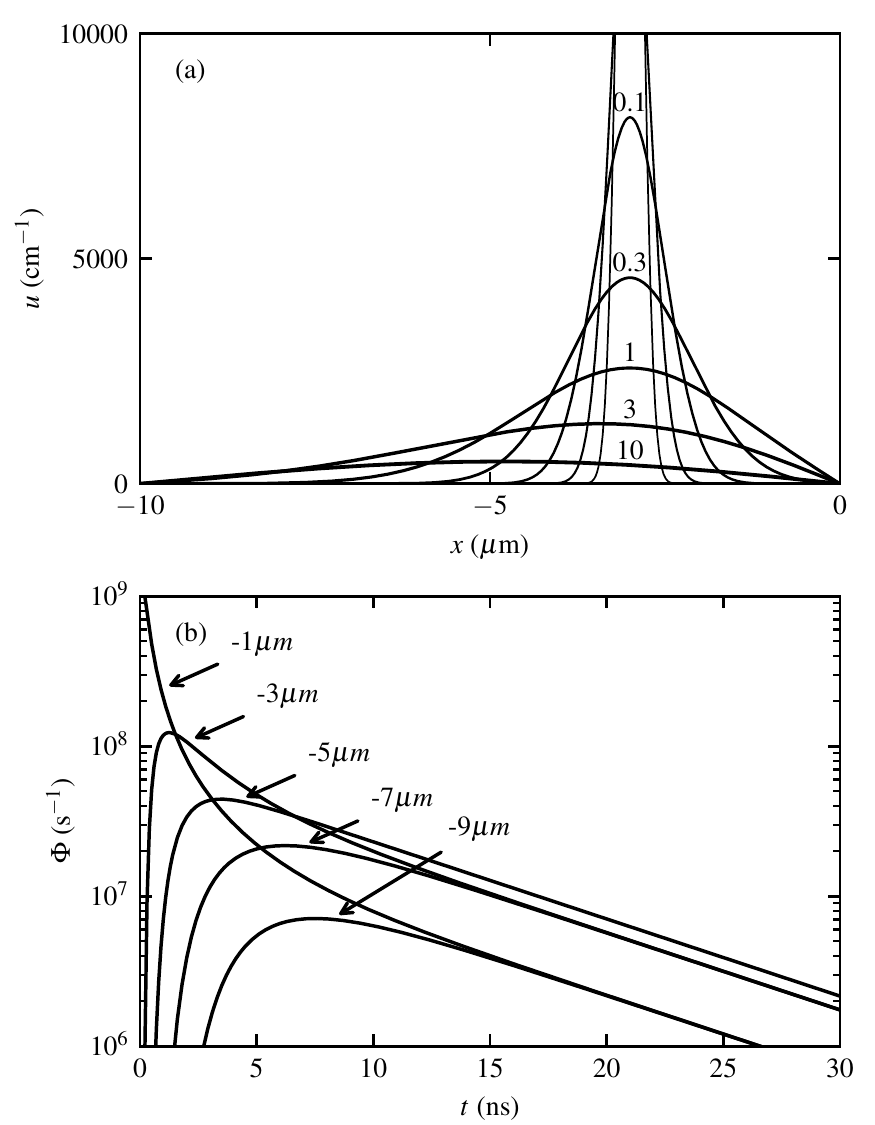}

\caption{Charge transport in undepleted area of \SI{10}{\micro\metre} for $V=\SI{80}{\volt}$; (a) time evolution of charge density along $x$ axis for a unitary charge deposited at $\xi_0=\SI{-3}{\micro\metre}$, with time indicated by labels (in \si{\nano\second}); note that at the boundaries, recombination and drift are relatively fast, resulting in a low carrier density near the boundaries; (b) shows the time evolution of the carrier flux from the undepleted to the depleted region for a range of initial positions (indicated by labels). For a \SI{10}{\micro\metre} undepleted region, the charge diffusion time is in the order of tens of \si{\nano\second}.}
\label{fig7}
\end{figure}

Fig.~\ref{fig7} shows in a first approximation the response of a relatively thin (\SI{10}{\micro\metre}) undepleted region of a partially depleted sensor ($V=\SI{80}{\volt}$). A full model would account for the coupling at the depletion boundary ($x=0$) of the differential equations on the two domains. However, this model already allows us to draw some initial conclusions on the signals from the undepleted region. Fig.~\ref{fig7}~(a) shows the evolution of the transient charge distribution ass a function of initial position $\xi$. 

The carrier flux entering the readout ASIC is given by the left to right flux through boundary $x=0$ at time $t$:
\begin{multline}
    \Phi(\xi,t)=D \pdv{g(x,\xi,t)}{x}\bigg|_{x \to 0} =\\
    = \frac{2 \pi D}{U^2}  \e^{-\gamma t} \sum_{n=1}^{\infty}  n \e^{-\frac{D \pi ^2 n^2 t}{U^2}} \sin \left(\frac{n \pi \xi }{U}\right) \label{eq41}
\end{multline}
Fig.~\ref{fig7}~(b) shows the coresponding transient charge flowing into the depleted region. For a relatively thin undepleted region, the minority carriers drift through the depletion boundary within tens of nanoseconds.

Total flux diffusing through boundary $x=0$ into the depleted region:
\begin{align}
    \Phi_t(0)=\int_{t=0}^{\infty} \Phi(t)
    =2 \pi D \sum_{n=1}^{\infty} \frac{n \sin(\frac{n \pi \xi}{U})}{n^2 D \pi^2+\gamma U^2} \label{eq42}
\end{align}
Note that the recombination rate $\gamma$ is now tied up inside the summation terms. For low recombination rates $\gamma U^2 \ll D \pi^2$:
\begin{align}
        \lim_{\gamma U^2 \ll D \pi^2} \Phi_t(0) = \frac{2}{\pi} \sum_{n=1}^{\infty} \frac{\sin(\frac{n \pi \xi}{U})}{n}=1-\frac{\xi}{U} \label{eq43}
\end{align}
reflecting charge loss through the front plane.

\section{Applications}
\subsection{X-ray Photon Detection}
\subsubsection{Transient Charge Clouds}
\begin{figure}[!t]
\centering
\includegraphics[width=\columnwidth]{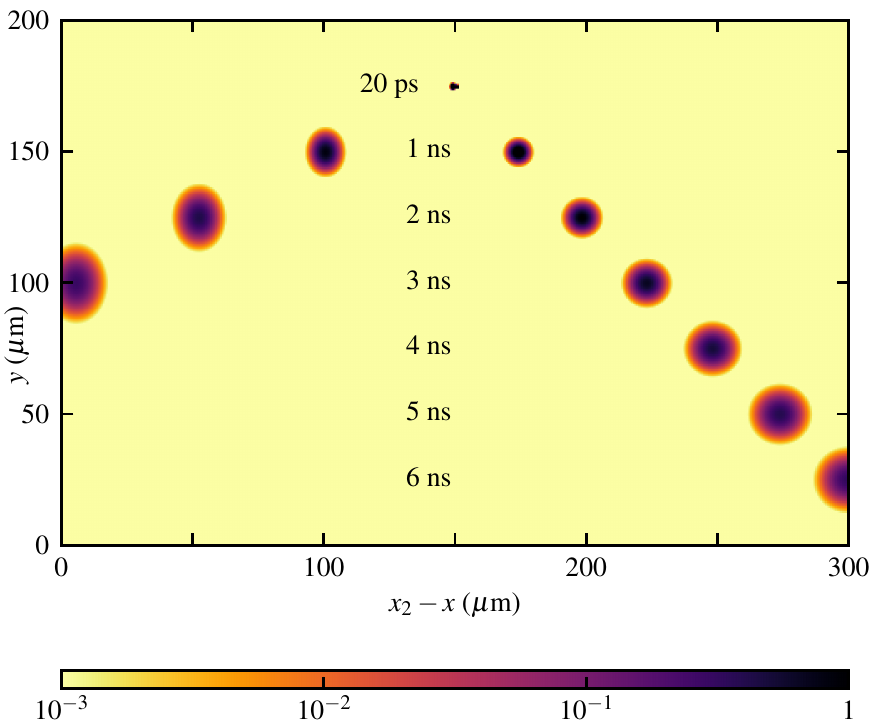}

\caption{
Charge drift for a single photon detected in the middle of a \SI{300}{\micro\metre} thick sensor ($x_2-x=\SI{250}{\micro\metre}$) with a bias voltage of \SI{200}{\volt}, after time intervals from \SIrange{2}{14}{\nano\second}. Left: electrons, right: holes. Figure shown on logarithmic scale over 3 orders of magnitude, with individual tracks scaled by the same amount. Note the extra stretching (holes) and compression (electrons) in the $x$ direction due to the electric field gradient.}
\label{fig8}
\end{figure}

Using the Green's function of the saturation velocity model (Eq.~\ref{eq32} and Eq.~\ref{eq35} inserted in Eq.~\ref{eq38}), we calculate the transient carrier distribution, project it on the $xy$ plane, and show the result in Fig.~\ref{fig8} for a typical sensor following a unit detection event in the middle of the \SI{300}{\micro\metre} sensor. The result is similar to Monte Carlo simulations, however, it is orders of magnitude faster.

\subsubsection{Transient Current}
The instantaneous current $i$ induced in one electrode due to the movement of one charge carrier is given by Ramo's theorem \cite{ramo1939currents} which requires taking into account the weighting potential \cite{riegler2016electric}. Note that the weighting potential is unrelated to the biasing or doping of the detector and can not be solved analytically in pixel detectors \cite{riegler2014point}. In pixel sensors the weighting potential decreases quickly away from a pixel readout pad, thus we can assume in a first approximation that the weighting field is $V\approx 0$ everywhere except on the current pixel readout pad, where $V\approx\SI{1}{\volt}$.

The expected current (in the statistical sense, as average of currents of many single carriers) for a single charge carrier through a boundary $x$ can be calculated easily from $g(x,\xi,t)$ (Eq.~\ref{eq38}):
\begin{align} \label{eq44}
    I(x,\xi,t)=e \left(v g - D \pdv{g}{x}\right)
    =e \left(v+D\frac{x-x_c}{\sigma^2 }\right) g(x,\xi,t) 
\end{align}

A typical detection event generates hundreds or thousands of carrier pairs, with currents approaching asymptotically the distribution shown in Eq.~\ref{eq44}, multiplied by the number of carriers. The current density resulting from $N$ carriers is: 
\begin{align} \label{eq45}
    J(x,y,z,t) =N I(x,\xi,t) w(y,z,t) 
\end{align}

\begin{figure}[!t]
\centering
\includegraphics[width=\columnwidth]{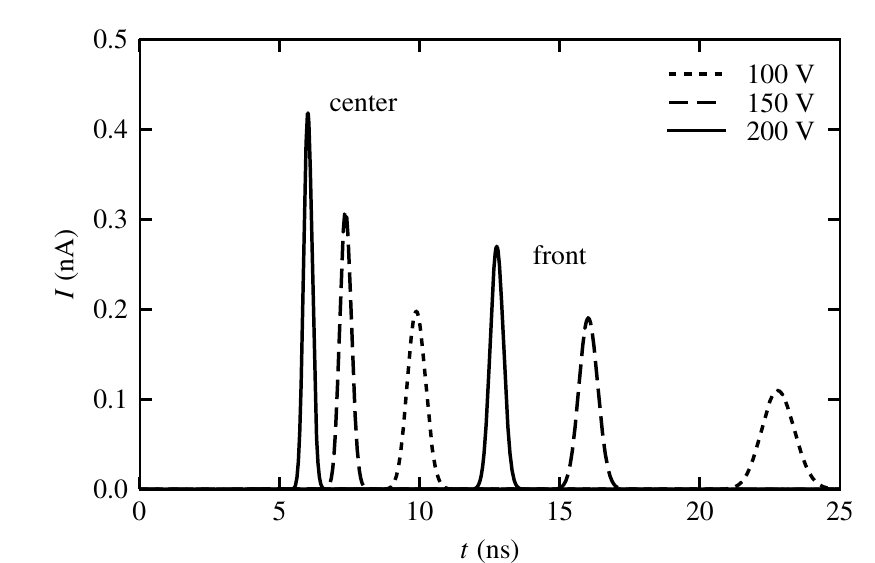}

\caption{Expected transient currents induced by minority carriers in a typical \SI{300}{\micro\metre}, n-type Si sensor as a function of time. The solid lines indicates a bias voltage of \SI{200}{\volt}, dashed lines correspond to \SI{150}{\volt} and dotted lines to \SI{100}{\volt}. The group of peaks at \SIrange{5}{11}{\nano\second} describe a carrier with initial position at the center of the sensor, while the group of peaks at \SIrange{12}{25}{\nano\second} correspond to an initial position at the front of the sensor. The drift time increases rapidly with lower bias and initial position farther away from the back plane.}
\label{fig9}
\end{figure}

In Fig.~\ref{fig9} we show the current from a single carrier in a typical sensor with initial position either at the front or in the middle of the \SI{300}{\micro\metre} sensor for 3 biasing conditions (\SI{100}{V}, \SI{150}{V} and \SI{200}{V}). The drift time increases rapidly with lower bias and initial position closer to the sensor front plane.

\subsubsection{Charge Cloud Size and Charge Sharing}
Assuming a square pixel in the $yz$ plane with pitch $L$ and center at $(0,0)$ collecting charge from a single carrier with initial position $(\xi_0,\upsilon_0,\zeta_0)$, the current entering the pixel at time $t$ is obtained by integrating $w(y,z,t)$ between the pixel boundaries, resulting in:
\begin{align} \label{eq46}
    I_1(t)=\frac{I(x,\xi_0,t)}{4} \erf\left(\frac{y}{\sqrt{2 D t}}\right)\Bigg |_{\upsilon_0-\frac{L}{2}}^{\upsilon_0+\frac{L}{2}} \erf\left(\frac{z}{\sqrt{2 D t}}\right)\Bigg |_{\zeta_0-\frac{L}{2}}^{\zeta_0+\frac{L}{2}}
\end{align}
Eq.~\ref{eq46} can be easily and efficiently extended to 2D pixel arrays.

\begin{figure}[!t]
\centering
\includegraphics[width=\columnwidth]{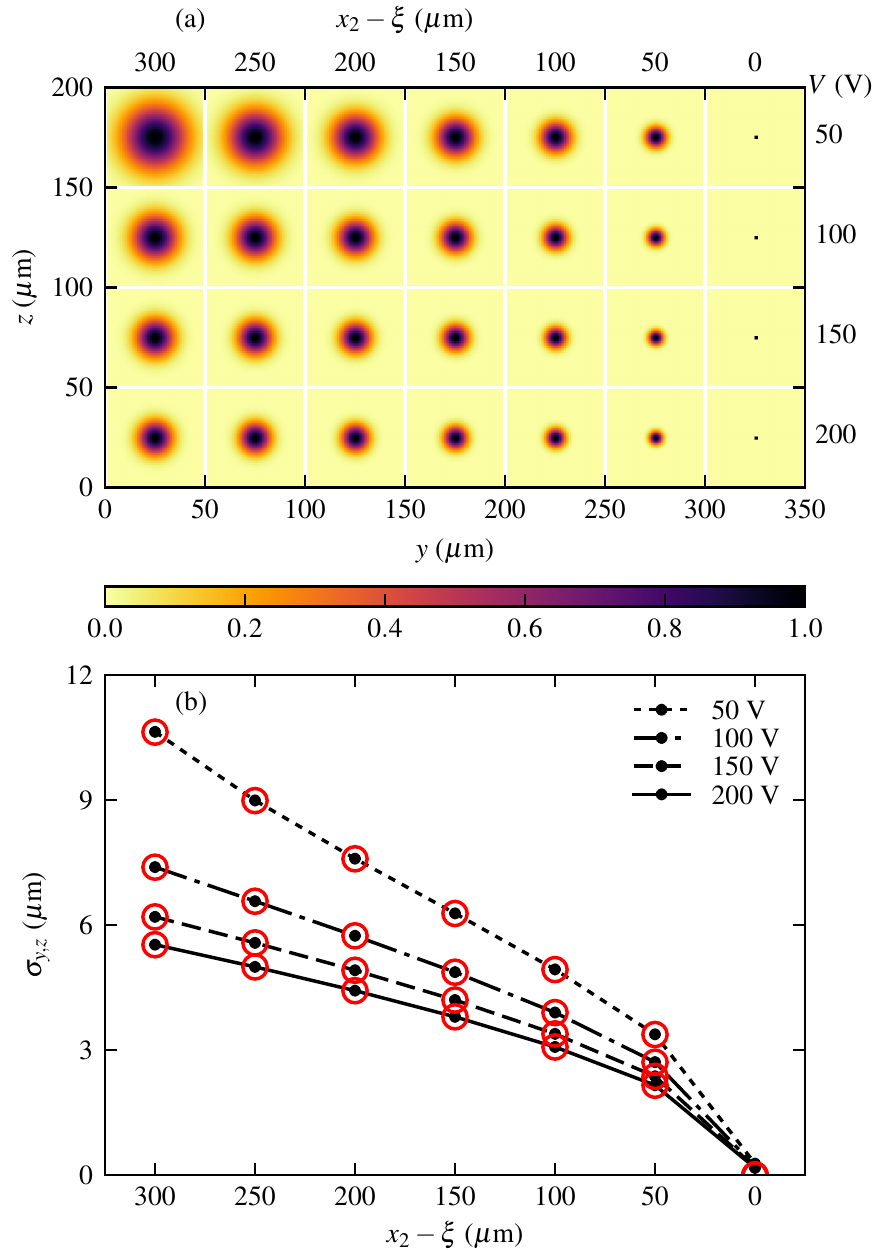}

\caption{
Charge clouds for single photons detected at different bias voltages (\SI{50}{\volt}, \SI{100}{\volt}, \SI{150}{\volt}, \SI{200}{\volt}) and different depths (from front surface at $x_2-\xi=\SI{300}{\micro\metre}$ to back surface at $x_2-\xi=0$), obtained by numerical integration of Eq.~\ref{eq46},~\ref{eq47}; (a) shows the lateral diffusion in the $y$ and $z$ directions; charge sharing increases rapidly with increasing distance from the back surface and with decreasing bias voltage; all clouds contain the same total charge, however, they are shown normalized to 1 in the center to emphasize the spatial charge sharing; For reference, white grid lines indicate edges of pixels with a pitch of \SI{50}{\micro\metre}; (b) shows the corresponding cloud sizes $\sigma_{y,z}$ (black dots); lower bias voltages and larger distances between the readout plane and interaction point ($x_2-\xi$) lead to increased charge cloud sizes.
}
\label{fig10}
\end{figure}

The total charge collected in a pixel can be calculated by integrating Eq.~\ref{eq46} numerically:
\begin{align} \label{eq47}
    Q_1=N \int_0^\infty I_1(t) dt
\end{align}
In Fig.~\ref{fig10}, black dots indicate results of the numerical integration (Eq.~\ref{eq46},~\ref{eq47}) for the typical sensor and a range of initial positions and biasing conditions.

\subsection{Relativistic Charged Particle Detection}
\subsubsection{Charge Cloud Size}
\begin{figure}[!t]
\centering
\includegraphics[width=\columnwidth]{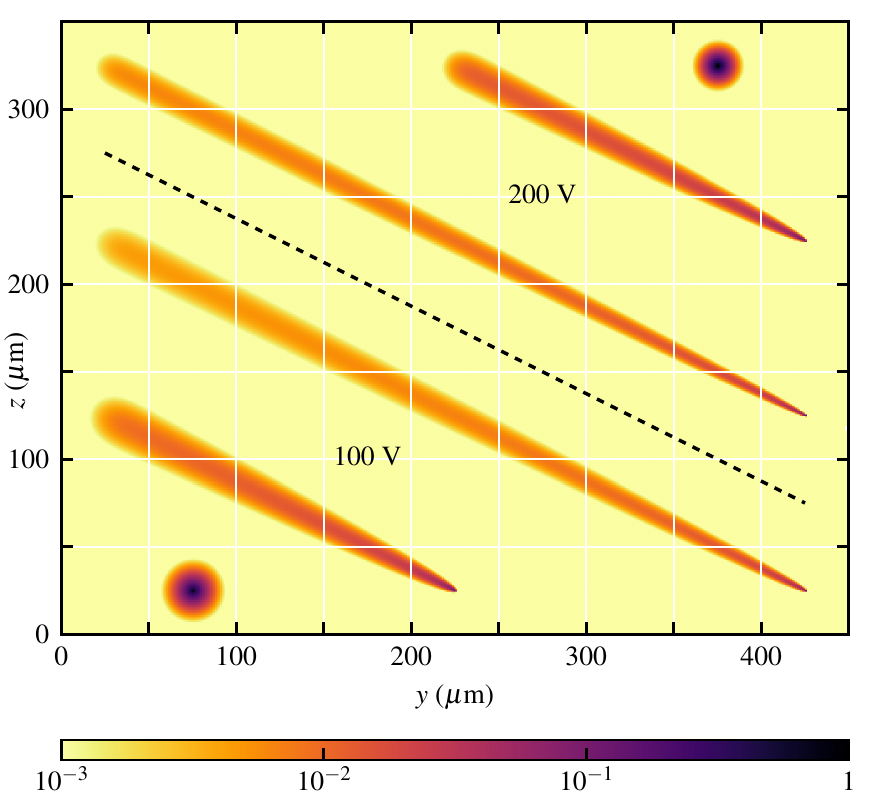}

\caption{Charge clouds for single relativistic electron tracks with three different inclinations ($\theta$ incident at \SI{90}{\degree}, \SI{53}{\degree} and \SI{34}{\degree} on the sensor plane) and 2 different bias voltages (\SI{200}{\volt} top right, \SI{100}{\volt} bottom left), assuming a typical \SI{300}{\micro\metre} thick sensor, n-type Si. Figure shown on logarithmic scale over 3 orders of magnitude, with individual tracks scaled by the same amount. For reference, white grid lines indicate edges of pixels with a pitch of \SI{50}{\micro\metre}.}
\label{fig11}
\end{figure}

In Fig.~\ref{fig11} we show charge sharing profiles for relativistic electron tracks for two different bias settings (\SI{200}{\volt} and \SI{100}{\volt}) and three different incidence angles (\SI{90}{\degree}, \SI{53}{\degree} and \SI{34}{\degree}).  For an application in tracking relativistic electron beams see \cite{blaj20163d}.

\section{Conclusion}

We present for the first time analytical solutions for fast and accurate calculation of transient carrier densities in \mbox{p-n} junction sensors by solving the drift-diffusion-recombination equations for the minority and majority carriers in a variety of conditions: undepleted, depleted with linear velocity (carrier velocity proportional to electric field), depleted with saturation velocity (carrier velocity transitioning from proportional to electric field to velocity saturation), and overdepleted (carriers moving with saturation velocity). We also show that the diffusion equations in the linear velocity model and in the saturation velocity model are increasingly generalized forms of the simple diffusion equation. We subsequently obtain the corresponding Green's functions which allow describing any initial conditions with a simple convolution.

Previously, analytical solutions were only available for simple diffusion (in the absence of drift velocity gradients). In practice, drift-diffusion is often simulated numerically with Monte Carlo simulations, finite elements simulations (including TCAD tools), or simple assumptions. Comparing our analytical drift-diffusion solutions with Monte Carlo simulations and TCAD simulations (using industry standard Synopsys Sentaurus) shows good agreement. 

We deduce equations for the transient behaviour of (1)~charge clouds resulting from detection of x-ray and gamma-ray photons at different depths and (2)~detection of relativistic charged particles and resulting tracks.
We illustrate the results for a typical silicon sensor (n-type, \SI{300}{\micro\metre} thick, resistivity$\rho=\SI{10}{\kilo\ohm\centi\metre}$), however, the analytical equations can be extended to any reverse-biased \mbox{p-n} junction pixel or strip sensor.

The transient charge cloud evolution can be used to describe the behaviour of timing pixel detectors. In particular, "time of arrival" and "time over threshold" measurements (defined as time until the transient signal exceeds a set threshold, and time elapsed until the transient signal returns under the set threshold) depend on threshold setting, photon energy, bias voltage, pixel geometry, 3D detection position (with subpixel accuracy); see holistic approaches to model these effects in Timepix for photons \cite{jakubek2011precise}, pions \cite{akiba2012charged}, protons and carbon ions \cite{hartmann2014distortion}. Timepix3 \cite{poikela2014timepix3} and tPix \cite{markovic2016design} have increasing "time of arrival" resolutions of \SI{1560}{\pico\second} and \SI{100}{\pico\second}, respectively.

Appropriate integration of the transient signals provides an accurate description of charge sharing for any pixel or strip detector for different conditions (interaction positions, track orientations, subpixel position, bias voltage, and pixel geometry) for photons and relativistic charged particle tracks. This can be used in extracting the 3D+T (x,y,z,time) of photon interactions (with subpixel resolution and accurate depth information) as well as the 4D+T (x,y,\texttheta,\textphi) track equation for relativistic charged particles.

\appendices
\section{Diffusion in Gradient Fields \label{appendix1}}
To separate diffusion from drift, we introduce a coordinate system change from $x$ to the position $s$ on the characteristic curve $x_c(\xi,t)$ \cite{logan2008introduction}:
\begin{align} \label{eq48}
    s=x-x_c
\end{align}
For a normal distribution, positions $x_c \mp \sigma$ correspond to locations which separate fractions $\left[1-\erf\left(\pm \sqrt{1/2}\right)\right]/2$ of the charge carriers (i.e., where $u_{xx}=0$). In this coordinate system, the evolution of the average standard deviation of the charge cloud $\sigma$ is described by:
\begin{align} \label{eq49}
    \dv{\sigma^2 }{t}=\pdv{\sigma^2 }{t}+\pdv{\sigma^2 }{s}\dv{s}{t}
\end{align}
where $\sigma$ is a function of $t$:
\begin{align} \label{eq50}
    \dv{\sigma^2 }{t}=2 \sigma \dv{\sigma}{t}
\end{align}
with the definition of $\sigma$ above, diffusion leads to: 
\begin{align} \label{eq51}
    \pdv{\sigma^2 }{t}=\pm 2 D
\end{align}
and $\sigma$ is identical with its position in coordinate $s$:
\begin{align} \label{eq52}
    \pdv{\sigma^2 }{s}=2 \sigma
\end{align}

We can also estimate: 
\begin{multline}
    \dv{s}{t}=\dv{(x-x_c)}{t}=v(x)-v(x_c)=v(s+x_c)-v(x_c) \label{eq53}
\end{multline}

Substituting Eq.~\ref{eq50}-\ref{eq53} in Eq.~\ref{eq49} and reducing by $2 \sigma$ ($\sigma>0$ for $t>0$) results in:
\begin{align} \label{eq54}
    \dv{\sigma}{t}=\frac{D}{\sigma}+v(\sigma+x_c)-v(x_c)
\end{align}

\section{Diffusion in Linear Velocity Model\label{appendix2}}
With $v(x)=b x$:
\begin{align} \label{eq55}
    v(\sigma+x_c)-v(\sigma)= b (\sigma+x_c)-b x_c = b \sigma
\end{align}
and substituting in Eq.~\ref{eq54} results in an ordinary differential equation:
\begin{align} \label{eq56}
    \dv{\sigma}{t}=\frac{D}{\sigma}+ b\sigma
\end{align}
with solution \cite{wolfram2017mathematica}:
\begin{align} \label{eq57}
    \sigma(t)=\sqrt{\frac{D}{b} \left(\e^{2 b t}-1\right)}
\end{align}
leading, with Eq.~\ref{eq24}, to a Green's function:
\begin{align} \label{eq58}
    g(x,\xi,t)=\frac{\exp \left(-\frac{ \left(x-\xi \e^{b t}\right)^2}{2 \frac{D}{b} \left(\e^{2 b t}-1\right)}\right)}{\sqrt{2 \pi} \sqrt{\frac{D}{b} \left(\e^{2 b t}-1\right)}}
\end{align}
which is an analytical solution for the partial differential equation. Note that the same result can be obtained \cite{polyanin2004handbook} by substituting $z=x+C \e^{-bt}$ and showing that Eq.~\ref{eq14} is reduced to an ordinary differential equation $a w(z)'' - b z w(z)' -b w = 0$ which is straightforward to solve and yields the same solution.

\section{Diffusion with Saturation Velocity\label{appendix3}}
We can expand Eq.~\ref{eq53} in Taylor series:
\begin{align} \label{eq59}
    \dv{s}{t}=v(\sigma+x_c)-v(\sigma)=\sum_{n=1}^\infty v^{(n)}(x_c) \frac{s^n}{n!} 
\end{align}
Substituting the actual saturation velocity formula (Eq.~\ref{eq8}), noting that $D \ll b d^2$ thus $\sigma \ll x_c$, and keeping only the first term, after some cancellation:
\begin{align} \label{eq60}
    \dv{s}{t}=v(\sigma+x_c)-v(\sigma) \approx \dv{v(x_c)}{x} \sigma = \frac{b v_s^2}{(v_s+b x_c)^2} \sigma
\end{align}
which results in an ordinary differential equation for diffusion:
\begin{align} \label{eq61}
    \dv{\sigma}{t}=\frac{D}{\sigma} + \frac{b v_s^2}{(v_s+b x_c(\xi,t))^2} \sigma
\end{align}

Substituting $x_c$ from Eq.~\ref{eq32} and solving analytically \cite{wolfram2017mathematica} the ordinary differential equation (Eq.~\ref{eq61}) with initial condition $\sigma(0)=0$, we obtain:
\begin{multline} \label{eq62}
    \sigma(\xi,t)=\\
    \frac{\sqrt{
    \frac{b D x_c^2}{v_s^2} \left(2 b t + 4 \ln(\frac{x_c}{\xi})\right)
    + \frac{6 D x_c}{v_s}(\frac{x_c}{\xi}-1) +\frac{D}{b}(\frac{x_c^2}{\xi^2}-1)
    }}{1+\frac{b x_c}{v_s}}
\end{multline}
which is a function of both $t$ and $\xi$ in the saturation velocity model; $x_c$ from Eq.~\ref{eq32}.

It is interesting to note that the series in Eq.~\ref{eq59} using $v(x)$ from Eq.~\ref{eq8} can be written as:
\begin{align} \label{eq63}
    \dv{s}{t}=-\frac{v_s^2}{v_s+b x_c} \sum_{n=1}^{\infty}\left(\frac{-\sigma}{\frac{v_s}{b}+x_c}\right)^n = \frac{v_s^2}{v_s+b x_c} \frac{\sigma}{\sigma + x_c + \frac{v_s}{b}}
\end{align}
with a corresponding ordinary differential equation:
\begin{align} \label{eq64}
    \dv{\sigma}{t}=\frac{D}{\sigma} + \frac{v_s^2}{v_s+b x_c(\xi,t)} \frac{\sigma}{\sigma + x_c(\xi,t) + \frac{v_s}{b}} 
\end{align}
which could be useful in finite elements analysis. However, this equation is difficult to solve analytically ($x_c$ depends on time $t$).

\section*{Acknowledgment}
Use of the Linac Coherent Light Source (LCLS), SLAC National Accelerator Laboratory, is supported by the U.S. Department of Energy, Office of Science, Office of Basic Energy Sciences under Contract No. DE-AC02-76SF00515.

The authors are grateful to C.~Genes (Max Plank Institute for the Science of Light, Erlangen, Germany) and A.~Dragone, C.~Stan and C.-E.~Chang (SLAC National Accelerator Laboratory, Menlo Park, California) for many stimulating discussions.

We applied the SDC approach for the sequence of authors \cite{tscharntke2007author}. Statement of authorship: conception, G.~Blaj; analytical methods, G.~Blaj; simulations, G.~Blaj and J.~Segal; drafting the manuscript, G.~Blaj; revising the manuscript: all authors.

\ifCLASSOPTIONcaptionsoff
  \newpage
\fi

\bibliographystyle{IEEEtran}
\bibliography{IEEEabrv,main}
\end{document}